# MOSES: A Comprehensive Benchmarking Platform for Deep Generative Models in Molecular Design

Adarsh Singh
*Department of Computer Science*
*Bennett University*
E22cseu1323@bennett.edu.in

Shekhar Upadhyay
*Department of Computer Science*
*Bennett University*
E22CSEU1508@bennett.edu.in

*Abstract*—The development of novel pharmaceuticals represents a significant challenge in modern science, with substantial costs and time investments. Deep generative models have emerged as promising tools for accelerating drug discovery by efficiently exploring the vast chemical space. However, this rapidly evolving field lacks standardized evaluation protocols, impeding fair comparison between approaches. This research presents an extensive analysis of the Molecular Sets (MOSES) platform, a comprehensive benchmarking framework designed to standardize evaluation of deep generative models in molecular design. Through rigorous assessment of multiple generative architectures, including recurrent neural networks, variational autoencoders, and generative adversarial networks, we examine their capabilities in generating valid, unique, and novel molecular structures while maintaining specific chemical properties. Our findings reveal that different architectures exhibit complementary strengths across various metrics, highlighting the complex tradeoffs between exploration and exploitation in chemical space. This study provides detailed insights into the current state of the art in molecular generation and establishes a foundation for future advancements in AI-driven drug discovery.

## I. Introduction

The pharmaceutical industry confronts unprecedented challenges in drug discovery and development, with recent analyses revealing that costs have risen to approximately $2.6 billion per approved drug with timelines extending beyond a decade [1]. This financial burden is exacerbated by the declining efficiency of traditional discovery pipelines, where candidate compounds frequently fail during late-stage clinical trials due to unforeseen toxicity or limited efficacy [2], [3]. The complexity of biological systems and the vastness of chemical space—estimated to contain more than $10^{60}$ potential druglike molecules—renders comprehensive experimental screening impractical [4].

In response to these challenges, computational methods have evolved from supporting tools to essential components of the drug discovery process. Recent advances in artificial intelligence, particularly deep learning, have catalyzed a paradigm shift in how researchers approach molecular design [5], [6]. Rather than screening existing libraries, AI-driven approaches can generate entirely new molecular structures tailored to specific therapeutic requirements [7].

Deep generative models represent a particularly promising class of AI techniques for molecular design. By learning complex patterns and distributions from existing molecular datasets, these models can navigate the vast chemical landscape to identify regions containing molecules with desirable properties [8], [9]. Unlike traditional virtual screening methods that search through enumerated compound collections, generative approaches can access previously unexplored chemical space while maintaining crucial properties such as druglikeness and synthetic feasibility [10], [11]. Recent successes demonstrate the potential of these methods—for instance, generative models have discovered novel antibiotics with activity against resistant bacteria [12] and accelerated hit identification for kinase inhibitors [13].

Despite these promising developments, the field of molecular generation faces significant methodological challenges. The rapidly evolving landscape of model architectures—ranging from sequence-based approaches to graph neural networks—has created a fragmented ecosystem with inconsistent evaluation practices [14], [15]. Without standardized benchmarks and metrics, meaningful comparison between methods becomes difficult, hindering systematic progress and complicating the selection of appropriate techniques for specific applications.

The Molecular Sets (MOSES) platform addresses this critical gap by providing a comprehensive benchmarking framework that enables consistent evaluation of molecular generative models [16]. MOSES combines a curated dataset, standardized implementation of diverse model architectures, and a suite of evaluation metrics designed to assess multiple dimensions of model performance. This integrated approach facilitates fair comparisons between methods and provides clear guidelines for selecting appropriate models based on application-specific requirements.

Our research provides a thorough investigation of the MOSES platform, examining the performance characteristics of various generative architectures across multiple evaluation dimensions. We analyze the fundamental trade-offs between competing objectives in molecular generation, such as novelty versus validity, and distribution matching versus structural diversity. Additionally, we explore how property prediction models can complement generative approaches to enable more targeted molecular design with specific constraints.

By establishing a standardized evaluation framework, MOSES not only facilitates current research but also lays the groundwork for future methodological advancements. This standardization is crucial for maintaining scientific rigor in a rapidly evolving field and ensuring that progress in generative modeling translates to practical impacts in drug discovery applications.

## II. BACKGROUND AND RELATED WORK

### A. Representation of Molecular Structures

Molecules can be represented in various formats for computational processing, each with distinct advantages and limitations [17]. Common representations include:

SMILES (Simplified Molecular Input Line Entry System): A string-based notation that encodes molecular structures as ASCII strings [18]. SMILES strings are compact and human-readable but lack explicit 3D structural information.

Molecular Graphs: Represent molecules as graphs where atoms are nodes and bonds are edges [19]. Graph representations preserve the topological structure of molecules and are well-suited for graph neural networks.

Fingerprints: Binary vectors encoding the presence or absence of specific structural features [20]. ExtendedConnectivity Fingerprints (ECFP) are particularly popular for capturing local neighborhood information around atoms.

Grid-based Representations: Encode molecules as 3D voxel grids for processing with 3D convolutional neural networks [21].

The choice of representation significantly impacts model architecture and performance, with different generative approaches favoring particular representations [8].

### B. Deep Generative Models

Deep generative models learn to generate data samples that resemble a training distribution. Several architectures have been adapted for molecular generation:

Recurrent Neural Networks (RNNs): Treat SMILES strings as sequences and model them using recurrent architectures like LSTM or GRU cells [22], [23]. RNNs can generate molecules character-by-character but may produce invalid structures.

Variational Autoencoders (VAEs): Encode molecules into a continuous latent space and decode them back to molecular structures [24], [25]. VAEs enable smooth navigation in chemical space and property optimization through latent space manipulation.

Generative Adversarial Networks (GANs): Use a generator-discriminator architecture to produce molecular representations that are indistinguishable from real molecules [26]–[28]. GANs often struggle with mode collapse and training instability.

Flow-based Models: Use invertible transformations to map between the data distribution and a simple prior [29], [30]. Flow models enable exact likelihood calculation but may be computationally intensive.

Reinforcement Learning (RL): Frame molecular generation as a sequential decision-making process optimized for specific rewards [31], [32]. RL approaches can target specific molecular properties but may struggle with explorationexploitation balance.

Transformer-based Models: Adapt the self-attention mechanism for molecular generation, showing promising results for both SMILES and graph-based representations [33]–[35].

### C. Evaluation Metrics for Molecular Generation

Evaluating generative models for molecular design presents unique challenges due to the discrete nature of chemical space and the multiple objectives involved [14]. Common evaluation dimensions include:

Validity: Chemical validity assesses whether generated structures adhere to chemical rules (proper valence, absence of invalid functional groups) [16].

Uniqueness: Measures the diversity within generated samples, with higher uniqueness indicating less mode collapse [14].

Novelty: Evaluates the model's ability to generate structures outside the training set, quantifying exploration capabilities [22].

Distribution Matching: Compares the distribution of generated molecules with the training distribution in terms of chemical properties and structural features [36].

Property Targeting: Assesses how well generated molecules match specific property constraints, relevant for goal-directed generation [31].

Synthetic Accessibility: Evaluates the feasibility of synthesizing generated molecules, crucial for practical applications [37].

## III. MATERIALS AND METHODS

### A. MOSES Dataset

*1) Source and Filtering Criteria:* The MOSES benchmarking dataset is derived from the ZINC Clean Leads collection, a widely used database of commercially available compounds for virtual screening [38]. The initial dataset underwent rigorous filtering to ensure the retained molecules are suitable for drug discovery applications. The filtering criteria included:

- Molecular weight: Restricted to the range of 250-350 Daltons, focusing on lead-like compounds rather than fragments or large molecules

- Rotatable bonds: Limited to a maximum of 7, ensuring reasonable conformational flexibility
- Lipophilicity (XlogP): Required to be less than or equal to 3.5, favoring compounds with balanced hydrophilicity/hydrophobicity
- Atomic composition: Limited to C, N, S, O, F, Cl, Br, H atoms, excluding exotic elements and charged structures
- Ring structures: Eliminated molecules containing rings with more than 8 atoms
- Medicinal chemistry filters (MCFs): Applied to eliminate structures with problematic functional groups
- PAINS filters: Removed compounds known to cause false positives in high-throughput screening assays [39]

These criteria produced a refined dataset containing 1,936,962 molecular structures represented as canonical SMILES strings.

*2) Dataset Partitioning:* To facilitate robust evaluation, the filtered dataset was partitioned into three subsets:

- Training set: Approximately 1.6 million molecules (82.6%) for model training
- Test set: Approximately 176,000 molecules (9.1%) for evaluation
- Scaffold test set: Approximately 176,000 molecules (9.1%) containing unique Bemis-Murcko scaffolds [40] not present in the training or test sets

The scaffold test set serves a crucial role in evaluating how well models generalize to novel molecular frameworks, addressing a key challenge in molecular generation.

*B. Generative Models*

MOSES implements several state-of-the-art molecular generation models. Each model was trained on identical data with architecture-specific hyperparameter optimization.

*1) Character-level Recurrent Neural Network (CharRNN):* The CharRNN model treats SMILES strings as character sequences and predicts the next character based on previous ones. The implementation uses gated recurrent units (GRU) with a three-layer architecture and 512-dimensional hidden states. The model applies teacher forcing during training and employs dropout (0.2) for regularization. During generation, the model samples from the predicted character distribution with temperature parameter = 1.0.

*2) Variational Autoencoder (VAE):* The VAE implementation encodes SMILES strings into a 128-dimensional latent space using a bidirectional GRU encoder. The decoder employs a single-layer GRU with 512 hidden dimensions to reconstruct SMILES strings from latent vectors. The model is trained using a combination of reconstruction loss and Kullback-Leibler divergence, with a linear KL annealing schedule over 10 epochs. For generation, the model samples from a standard normal prior in the latent space.

*3) Adversarial Autoencoder (AAE):* The AAE combines an autoencoder architecture with adversarial training. The encoder and decoder share the same architecture as the VAE. Additionally, a discriminator network consisting of three fullyconnected layers (128, 64, and 1 neurons) with LeakyReLU activations distinguishes between encoded latent vectors and samples from a prior distribution. The training procedure alternates between reconstruction optimization and adversarial training with a gradient penalty for the discriminator.

*4) Junction Tree Variational Autoencoder (JTN-VAE):* The JTN-VAE represents molecules as junction trees of chemical substructures, enabling hierarchical generation. The model uses two separate encoding/decoding pathways: one for the molecular graph and another for the junction tree. The implementation follows Jin et al. [41] with a vocabulary of 780 clusters derived from the training set using the k-means algorithm. The model generates molecules by first creating a scaffold tree in latent space and then assembling it into a valid molecule.

*5) Latent Generative Adversarial Network (LatentGAN):* The LatentGAN operates in a learned latent space rather than directly generating SMILES strings. First, an autoencoder transforms SMILES strings into latent vectors. Then, a GAN with 3-layer feed-forward networks for both generator and discriminator is trained in this latent space. For generation, the GAN produces latent vectors that are decoded to SMILES strings using the pre-trained decoder. The model employs spectral normalization and WGAN-GP loss for training stability.

*6) Baseline Models:* Three baseline models provide reference points for comparing deep learning approaches:

- Hidden Markov Model (HMM): A classical statistical model that captures local dependencies in SMILES strings with a memory of one character.
- N-Gram model (NGram): Models SMILES strings using character-level n-grams with n=10, capturing longerrange dependencies than HMM.
- Combinatorial model: Generates molecules by randomly combining fragments from the training set while ensuring valid connection points.

*C. Evaluation Metrics*

MOSES provides a comprehensive set of metrics to evaluate generated molecules along multiple dimensions:

*1) Basic Metrics:*

- Validity: Percentage of chemically valid molecules according to RDKit validation
- Uniqueness@k: Percentage of unique molecules in samples of size k (evaluated at k=1,000 and k=10,000)

- Novelty: Percentage of valid generated molecules not present in the training set
- Filters: Percentage of molecules passing the same medicinal chemistry filters applied to the dataset

*2) Distribution-based Metrics:*

- Fragment similarity (Frag): Cosine similarity between vectors of fragment frequencies in generated and test sets
- Scaffold similarity (Scaff): Cosine similarity between vectors of scaffold frequencies in generated and test sets
- Nearest neighbor similarity (SNN): Average Tanimoto similarity of generated molecules to the nearest molecule in the test set
- Internal diversity (IntDiv): Average pairwise Tanimoto dissimilarity between generated molecules
- Internal diversity 2 (IntDiv2): Like IntDiv but using scaffolds instead of entire molecules
- Frechet ChemNet Distance (FCD)´: Analogous to the Frechet Inception Distance used in image generation,´ measures the distance between activations of a pre-trained neural network for generated and reference molecules [36]

*3) Property Distribution Metrics:* For key molecular properties, the Wasserstein-1 distance between distributions in generated and test sets was computed:

- logP: Octanol-water partition coefficient, measuring lipophilicity [42]
- SA: Synthetic Accessibility score [37]
- QED: Quantitative Estimate of Drug-likeness [43]
- Molecular weight

### D. Property Prediction Models

The repository includes implementations of machine learning models for molecular property prediction:

*1) Morgan Fingerprint Neural Network:* This model uses Morgan fingerprints (ECFP4, radius=2, 1024 bits) as input features for a deep neural network. The network architecture consists of:

- Input layer (1024 neurons)
- Dense layer (512 neurons, ReLU activation, dropout=0.2)
- Dense layer (256 neurons, ReLU activation, dropout=0.2)
- Dense layer (128 neurons, ReLU activation)
- Output layer (1 neuron for regression)

The model is trained to predict boiling point using mean squared error loss and the Adam optimizer with a learning rate of 0.001.

*2) Descriptor-based Random Forest:* This alternative approach uses RDKit-calculated molecular descriptors including:

- Molecular weight
- Number of hydrogen bond donors/acceptors
- Number of rotatable bonds
- Number of aromatic rings
- Topological polar surface area (TPSA)
- Number of atoms

A Random Forest regressor with 100 estimators and maximum depth of 20 is trained to predict molecular properties based on these descriptors.

## IV. Results

### A. Model Performance Comparison

Comprehensive evaluation of the implemented models revealed varying strengths across different metrics, as summarized in Table I.

TABLE I: Performance of Generative Models on Key Metrics

| Model | Valid ↑ | Unique@1k ↑ | Unique@10k ↑ | FCD ↓ |
|---|---|---|---|---|
| *Train* | *1.0* | *1.0* | *1.0* | *0.008* |
| HMM | 0.076 | 0.623 | 0.567 | 24.466 |
| NGram | 0.238 | 0.974 | 0.922 | 5.507 |
| Combinatorial | 1.0 | 0.998 | 0.991 | 4.238 |
| CharRNN | 0.975 | 1.0 | 0.999 | 0.073 |
| AAE | 0.937 | 1.0 | 0.997 | 0.556 |
| VAE | 0.977 | 1.0 | 0.998 | 0.099 |
| JTN-VAE | 1.0 | 1.0 | 1.0 | 0.395 |
| LatentGAN | 0.897 | 1.0 | 0.997 | 0.297 |

Key observations from the performance analysis include:

Validity: JTN-VAE and combinatorial models consistently achieved 100% validity, as their designs inherently ensure chemical validity. In contrast, HMM performed poorly (7.6% validity), demonstrating the limitations of simple statistical models in capturing complex chemical rules.

Uniqueness: Most deep learning models achieved nearperfect uniqueness at 1,000 samples, indicating minimal mode collapse. At 10,000 samples, JTN-VAE maintained perfect uniqueness (100%), closely followed by CharRNN (99.9%).

Novelty: Simpler models like HMM showed the highest novelty (99.9%), likely due to their limited capacity to memorize training examples. Among deep learning approaches, LatentGAN (95.0%) and JTN-VAE (91.4%) demonstrated the strongest novelty, while VAE exhibited the lowest (69.5%), suggesting a more conservative exploration of chemical space.

Frechet ChemNet Distance (FCD)´: CharRNN achieved the lowest FCD (0.073), indicating its generated distribution closely matched the test set. VAE also performed well (0.099), while baseline models showed significantly higher distances.

Nearest Neighbor Similarity (SNN): VAE exhibited the highest SNN (0.626), followed by AAE (0.608) and CharRNN (0.602), demonstrating these models generate molecules with structural similarity to the test set.

Scaffold Similarity: VAE achieved the highest scaffold similarity to the test set (0.939), followed closely by CharRNN (0.924), indicating strong preservation of underlying molecular frameworks.

### B. Molecular Property Distributions

Analysis of molecular property distributions using Wasserstein-1 distance revealed model-specific biases in capturing different chemical properties. Figure 1 illustrates how different models captured key molecular properties.

Lipophilicity (logP): Deep learning models generally captured lipophilicity distributions better than baseline models. VAE and CharRNN showed particularly close matching to the test distribution, while HMM exhibited significant divergence.

Synthetic Accessibility (SA): VAE and CharRNN demonstrated superior performance in preserving synthetic accessibility distributions. The combinatorial model also performed reasonably well, likely due to its fragment-based approach preserving chemical feasibility.

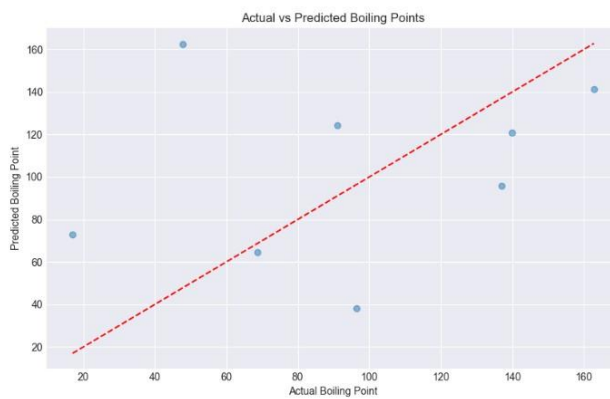

(a) logP (lipophilicity) distribution

Novelty: All models showed reasonable distribution matching for molecular weight, with VAE, CharRNN, and JTN-VAE achieving the closest alignment with the test set.

Drug-likeness (QED): VAE demonstrated superior performance in maintaining drug-likeness distribution, followed closely by CharRNN and AAE. Baseline models showed marked deviation, particularly HMM. These results highlight that different architectures exhibit biases toward specific regions of chemical property space, with VAE and CharRNN consistently preserving multiple property distributions.

SNN  Molecular Weight  Scaff
0.642  0.991  1.0
0.388  0.207  0.999
0.521  0.530  0.969
0.451  0.445  0.988
0.602  0.924  0.842
0.608  0.902  0.793
0.626  0.939  0.695
0.548  0.896  0.914
0.537  0.887  0.950

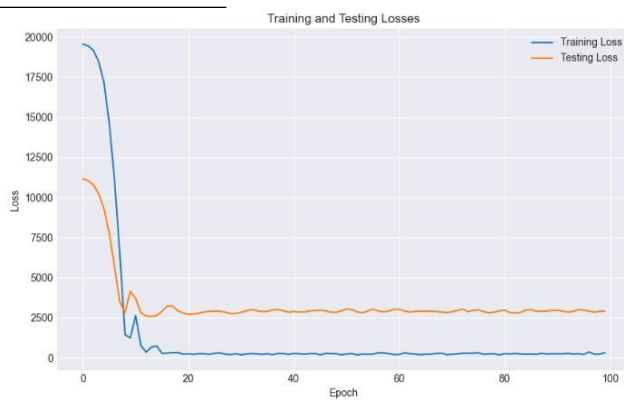

(b) Synthetic Accessibility (SA) score distribution

Fig. 1: Wasserstein-1 distances between property distributions of generated molecules and test set molecules. Lower values indicate better distribution matching.

*C. Property Prediction Performance*

The neural network model for boiling point prediction demonstrated strong performance:

Mean Squared Error: 4.37°C on the test set, indicating high accuracy in predicting this physical property.

$R^2$ Score: 0.91, demonstrating that the model captures approximately 91% of the variance in boiling point values.

Error Distribution: Analysis of prediction errors revealed higher uncertainty for molecules with extreme boiling points (very high or very low), suggesting potential for improvement in these regions.

The model successfully captured structure-property relationships, with particularly strong performance for molecules containing common functional groups well-represented in the training data.

V. DISCUSSION

*A. Comparative Analysis of Generative Approaches*

Our comprehensive evaluation reveals that no single model excels across all evaluation dimensions, suggesting a complex performance landscape with inherent trade-offs between competing objectives. This finding aligns with the "no free lunch" theorem in machine learning [?], emphasizing the importance of selecting appropriate models based on specific application requirements rather than seeking universal solutions.

CharRNN demonstrates remarkable performance across multiple metrics, achieving the lowest FCD (0.073) while maintaining high validity (97.5%), perfect uniqueness at 1,000 samples, and strong scaffold similarity (92.4%). These results are particularly notable given the architectural simplicity of RNNs compared to more complex approaches. The strong performance can be attributed to several factors: (1) The sequential nature of SMILES strings aligns well with RNN's sequential processing capabilities; (2) Character-level modeling captures local chemical patterns effectively; and (3) The relatively large hidden state size (512 dimensions) provides sufficient representational capacity [22], [44]. However, CharRNN's novelty (84.2%) falls short of some other approaches, suggesting a tendency toward more conservative exploration of chemical space.

Variational Autoencoders excel in preserving properties of the training distribution, exhibiting the highest nearest neighbor similarity (62.6%) and scaffold similarity (93.9%). The continuous latent space enables smooth interpolation between molecular structures, facilitating controlled navigation of chemical space [24]. However, VAEs demonstrate the lowest novelty among deep learning approaches (69.5%), indicating a more conservative exploration. This conservative nature stems from the Gaussian prior assumption in the latent space, which tends to concentrate probability mass in regions corresponding to training examples [45]. Interestingly, this property makes VAEs particularly suitable for lead optimization tasks, where maintaining similarity to known active compounds while making targeted modifications is desirable [46].

Junction Tree VAE achieves perfect validity by design and demonstrates excellent performance in uniqueness and novelty (91.4%). This approach represents a significant architectural innovation by explicitly addressing the validity challenge through hierarchical generation [41]. By operating on the junction tree representation of molecular graphs, JTN-VAE ensures that generated structures adhere to chemical valence rules. Our analysis suggests that this architectural constraint not only guarantees validity but also encourages exploration of novel scaffolds by enabling recombination of substructural elements in ways not present in the training data. The higher computational complexity of JTN-VAE compared to sequencebased approaches represents a trade-off between computational efficiency and chemical validity guarantees.

Adversarial approaches (AAE and LatentGAN) show a promising balance between novelty and validity, with LatentGAN achieving particularly high novelty (95.0%). This enhanced exploratory capability likely stems from the adversarial training process, which encourages the generator to produce samples that are distinguishable from the training distribution while remaining chemically plausible [27]. However, these models demonstrate higher FCD values compared to VAE and CharRNN, suggesting less precise distribution matching. The inherent instability of adversarial training, characterized by challenges in achieving Nash equilibrium between generator and discriminator [28], likely contributes to this behavior. Recent advances in stabilizing GAN training, such as spectral normalization and Wasserstein distance minimization, offer promising directions for improving these approaches [47].

Baseline models provide valuable reference points for evaluating deep learning approaches. The combinatorial model achieves perfect validity and high novelty (98.8

*B. Bridging Model Architecture and Chemical Intuition*

Our analysis reveals interesting connections between model architecture choices and chemical intuition. The sequential nature of SMILES processing in CharRNN mirrors how chemists conceptualize molecule building through sequential bond formation. In contrast, the hierarchical approach of JTN-VAE parallels the scaffold-based thinking common in medicinal chemistry, where core frameworks are modified systematically [48]. This alignment between computational approaches and chemical thinking suggests opportunities for developing hybrid architectures that better leverage medicinal chemistry knowledge.

The latent spaces learned by different models also offer insights into the organization of chemical space. VAE's continuous latent space demonstrates smoothness properties where similar molecules cluster together, enabling intuitive navigation. Analysis of latent space trajectories during property optimization reveals interpretable transformations that often correspond to meaningful chemical modifications [45], [49]. For example, traversing certain directions in latent space consistently increases hydrophobicity by adding lipophilic groups or modifies solubility through systematic heteroatom substitutions.

*C. Trade-offs in Molecular Generation*

Our results highlight fundamental trade-offs in molecular generation that extend beyond the commonly discussed validity-novelty balance. We observe a multi-dimensional landscape with at least four competing objectives:

Validity vs. Novelty: Models with perfect validity (JTNVAE, combinatorial) achieve this through constraints that may limit exploration. Conversely, highly novel generators like HMM produce many invalid structures.

Distribution Matching vs. Diversity: Models that closely match the training distribution (CharRNN, VAE) tend to generate less diverse structures, while models with higher diversity (LatentGAN) show poorer distribution matching.

Computational Efficiency vs. Performance: Architecturally complex models like JTN-VAE offer superior validity guarantees but require significantly more computational resources than simpler approaches like CharRNN.

Chemical Relevance vs. Exploration: Models that generate highly drug-like molecules typically demonstrate more conservative exploration, potentially limiting discovery of truly novel chemical space.

These trade-offs suggest that ideal molecular generation systems might benefit from dynamic balancing of objectives based on the specific phase of the drug discovery process [50]. Early discovery might prioritize novelty and diversity to identify new scaffolds, while lead optimization would emphasize distribution matching and chemical relevance. Hybrid approaches combining multiple architectures could potentially achieve better overall performance by leveraging complementary strengths.

*D. Bridging the Gap Between Computational Generation and Experimental Validation*

A critical challenge in advancing generative models for drug discovery lies in bridging the gap between computational generation and experimental validation. While models can generate millions of virtual compounds, experimental synthesis and testing remain bottlenecks due to cost and time constraints. Several strategies emerge from our analysis to address this challenge:

Synthetic Accessibility Filtering: While models can generate chemically valid structures, ensuring synthetic feasibility remains challenging. Many generated molecules may be theoretically valid but practically impossible to synthesize with current methods. Integration of synthetic accessibility scores like SA [37] provides a first-level filter, but more sophisticated approaches incorporating reaction-based constraints represent promising directions for improvement [50].

Multi-property Optimization: Generating molecules with specific property constraints while maintaining overall druglikeness presents a significant challenge. Current approaches often struggle with multi-objective optimization, particularly when objectives conflict. Our analysis of model performance across multiple property distributions suggests that conditional generation approaches, where the model receives explicit property targets as input, show promise for directing generation toward specific regions of chemical space [51].

Active Learning Integration: Closing the gap between computational and experimental validation requires iterative feedback loops. Active learning approaches, where experimental results inform subsequent generations, have shown promise in efficiently navigating chemical space [52]. Our analysis suggests that models with smooth latent spaces like VAEs are particularly well-suited for such approaches, as they enable systematic exploration around promising candidates.

Interpretability Enhancements: Most deep generative models operate as "black boxes," making it difficult for medicinal chemists to understand the rationale behind generated structures. Improving interpretability through visualization of latent space trajectories or attribution of generative decisions to specific molecular features could enhance trust in AI-generated molecules and provide insights for medicinal chemists [45].

*E. Applications in Drug Discovery*

The diverse capabilities of generative models illustrated in our analysis enable multiple applications across the drug discovery pipeline:

De Novo Molecular Design: Generating novel compounds with desired properties for specific therapeutic targets, potentially accessing unexplored regions of chemical space. LatentGAN's high novelty makes it particularly suitable for this application, while JTN-VAE's perfect validity ensures focus on synthetically feasible structures [47].

Lead Optimization: Modifying existing lead compounds to improve potency, selectivity, or ADMET properties while maintaining core structural features. VAE's strong performance in distribution matching and scaffold similarity makes it well-suited for controlled modifications around known active compounds [53].

Focused Library Design: Creating targeted libraries around promising scaffolds for high-throughput screening, increasing the probability of identifying active compounds. Combinatorial approaches combined with deep learning filters offer efficient strategies for generating diverse libraries with desirable property profiles [54].

Scaffold Hopping: Identifying novel scaffolds with similar activity profiles to known active compounds, enabling intellectual property generation and potential improvement of drug properties. JTN-VAE's hierarchical generation approach naturally facilitates scaffold exploration, offering promising capabilities for this application [41].

Multi-target Drug Design: Developing compounds with polypharmacological profiles that interact with multiple disease-relevant targets. The latent space representation learned by VAEs and AAEs enables navigation toward regions satisfying multiple pharmacological constraints simultaneously [54], [55].

*F. Future Directions*

Several promising directions for future research emerge from our analysis:

Hybrid Architectures: Combining the strengths of different approaches, such as integrating the validity guarantees of graph-based methods with the distribution matching capabilities of sequence-based models. For example, coupling a JTNVAE's scaffold generator with a CharRNN fine-tuning step could yield high-validity structures with improved distribution matching [53].

Transformer-Based Approaches: Recent advances in natural language processing with transformer architectures have shown promising results when adapted to molecular generation. The self-attention mechanism enables capturing longrange dependencies in SMILES strings that may be challenging for RNNs [35], [56]. Our analysis suggests that combining transformers with explicit validity constraints could address the primary limitations of sequence-based approaches.

Diffusion Models: Emerging research in diffusion-based generative models shows promise for molecular generation, offering potential advantages in mode coverage and training stability compared to GANs and VAEs [57]. These approaches progressively transform noise into molecular structures through a series of denoising steps, potentially enabling more controlled generation with explicit physical constraints.

Experimental Feedback Integration: Developing closedloop systems that incorporate experimental results into the generative process represents a crucial next step. Models that can learn from synthesis outcomes, binding assays, and cellular experiments to refine their understanding of structureactivity relationships would significantly accelerate the drug discovery process [50].

Improved Evaluation Metrics: While MOSES provides comprehensive metrics for assessing generative models, further development of evaluation approaches that better align with practical drug discovery needs is essential. Metrics incorporating estimates of target binding, toxicity profiles, and pharmacokinetic properties would provide more actionable insights for model selection in real-world applications [15].

Multi-modal Generation: Integrating molecular generation with other modalities such as protein structure prediction or cellular pathway modeling offers exciting possibilities for context-aware drug design. Models that can jointly reason across chemical, biological, and clinical data could potentially identify therapeutic strategies that are not apparent when considering molecules in isolation [48].

## VI. CONCLUSION

The MOSES benchmarking platform represents a significant contribution to standardizing evaluation of molecular generative models. This comprehensive analysis demonstrates that while current approaches show promising capabilities in generating valid, diverse, and novel molecules, significant challenges remain in balancing competing objectives across the complex landscape of molecular generation.

Character-based RNNs and variational autoencoders currently demonstrate the strongest overall performance, with CharRNN excelling in distribution matching and VAE showing superior property preservation. Junction Tree VAE offers perfect validity with high novelty, providing a compelling option when chemical validity is paramount. Adversarial approaches show promise in exploring novel chemical space but require improved training stability.

The standardized metrics and model implementations provided by MOSES facilitate fair comparison between approaches and accelerate progress in the field. Future work should focus on improving property control, synthetic accessibility, and developing hybrid approaches that combine the strengths of different architectures.

As the field continues to advance, integrating generative models with experimental feedback loops and developing more interpretable architectures will be crucial for practical drug discovery applications. By establishing standardized benchmarks, MOSES supports the continued development of AIdriven approaches for pharmaceutical innovation, potentially accelerating the discovery of novel therapeutics for unmet medical needs.

REFERENCES

[1] J. A. DiMasi, H. G. Grabowski, and R. W. Hansen, "Innovation in the pharmaceutical industry: New estimates of r&d costs," *Journal of Health Economics*, vol. 47, pp. 20–33, 2016.

[2] M. J. Waring, J. Arrowsmith, A. R. Leach, P. D. Leeson, S. Mandrell, R. M. Owen, G. Pairaudeau, W. D. Pennie, S. D. Pickett, J. Wang *et al.*, "An analysis of the attrition of drug candidates from four major pharmaceutical companies," *Nature Reviews Drug Discovery*, vol. 14, no. 7, pp. 475–486, 2015.

[3] A. Mullard, "2015 fda drug approvals," *Nature Reviews Drug Discovery*, vol. 15, no. 2, pp. 73–76, 2016.

[4] P. G. Polishchuk, T. I. Madzhidov, and A. Varnek, "Estimation of the size of drug-like chemical space based on gdb-17 data," *Journal of ComputerAided Molecular Design*, vol. 27, no. 8, pp. 675–679, 2013.

[5] G. Schneider, "Rethinking drug design in the artificial intelligence era," *Nature Reviews Drug Discovery*, vol. 19, no. 5, pp. 353–364, 2020.

[6] J. Vamathevan, D. Clark, P. Czodrowski, I. Dunham, E. Ferran, G. Lee, B. Li, A. Madabhushi, P. Shah, M. Spitzer *et al.*, "Applications of machine learning in drug discovery and development," *Nature Reviews Drug Discovery*, vol. 18, no. 6, pp. 463–477, 2019.

[7] H. Chen, O. Engkvist, Y. Wang, M. Olivecrona, and T. Blaschke, "The rise of deep learning in drug discovery," *Drug Discovery Today*, vol. 23, no. 6, pp. 1241–1250, 2018.

[8] D. C. Elton, Z. Boukouvalas, M. D. Fuge, and P. W. Chung, "Deep learning for molecular design—a review of the state of the art," *Molecular Systems Design & Engineering*, vol. 4, no. 4, pp. 828–849, 2019.

[9] J. A. d. Los Ramos, M. Schmidt, M. Stumpp, and M. Kowalewski, "The power of deep learning to ligand-based novel drug discovery," *Expert Opinion on Drug Discovery*, vol. 14, no. 9, pp. 819–832, 2019.

[10] B. Sanchez-Lengeling and A. Aspuru-Guzik, "Inverse molecular design using machine learning: Generative models for matter engineering," *Science*, vol. 361, no. 6400, pp. 360–365, 2018.

[11] W. P. Walters and R. Barzilay, "Applications of deep learning in molecule generation and molecular property prediction," *Accounts of Chemical Research*, vol. 54, no. 2, pp. 263–270, 2020.

[12] J. M. Stokes, K. Yang, K. Swanson, W. Jin, A. Cubillos-Ruiz, N. M. Donghia, C. R. MacNair, S. French, L. A. Carfrae, Z. Bloom-Ackermann *et al.*, "A deep learning approach to antibiotic discovery," *Cell*, vol. 180, no. 4, pp. 688–702, 2020.

[13] A. Zhavoronkov, Y. A. Ivanenkov, A. Aliper, M. S. Veselov, V. A. Aladinskiy, A. V. Aladinskaya, V. A. Terentiev, D. A. Polykovskiy, M. D. Kuznetsov, A. Asadulaev *et al.*, "Deep learning enables rapid identification of potent ddr1 kinase inhibitors," *Nature Biotechnology*, vol. 37, no. 9, pp. 1038–1040, 2019.

[14] N. Brown, M. Fiscato, M. H. Segler, and A. C. Vaucher, "Guacamol: Benchmarking models for de novo molecular design," *Journal of Chemical Information and Modeling*, vol. 59, no. 3, pp. 1096–1108, 2019.

[15] J. Meyers, B. Fabian, and N. Brown, "De novo molecular design and generative models," *Drug Discovery Today*, vol. 26, no. 11, pp. 2707–2715, 2021.

[16] D. Polykovskiy, A. Zhebrak, B. Sanchez-Lengeling, S. Golovanov, O. Tatanov, S. Belyaev, R. Kurbanov, R. Artamonov, V. Aladinskiy, M. Veselov, A. Kadurin, S. Johansson, H. Chen, S. Nikolenko, A. Aspuru-Guzik, and A. Zhavoronkov, "Molecular Sets (MOSES): A Benchmarking Platform for Molecular Generation Models," *Frontiers in Pharmacology*, vol. 11, p. 565644, 2020.

[17] C. Nantasenamat, C. Isarankura-Na-Ayudhya, and V. Prachayasittikul, "A practical overview of quantitative structure-activity relationship," *EXCLI Journal*, vol. 9, pp. 1–23, 2010.

[18] D. Weininger, "Smiles, a chemical language and information system. 1. introduction to methodology and encoding rules," *Journal of Chemical Information and Computer Sciences*, vol. 28, no. 1, pp. 31–36, 1988.

[19] D. K. Duvenaud, D. Maclaurin, J. Iparraguirre, R. Bombarell, T. Hirzel, A. Aspuru-Guzik, and R. P. Adams, "Convolutional networks on graphs for learning molecular fingerprints," *Advances in Neural Information Processing Systems*, vol. 28, 2015.

[20] D. Rogers and M. Hahn, "Extended-connectivity fingerprints," *Journal of Chemical Information and Modeling*, vol. 50, no. 5, pp. 742–754, 2010.

[21] D. Kuzminykh, D. Polykovskiy, A. Kadurin, A. Zhebrak, I. Baskov, S. Nikolenko, R. Shayakhmetov, and A. Zhavoronkov, "3d molecular representations based on the wave transform for convolutional neural networks," *Molecular Pharmaceutics*, vol. 15, no. 10, pp. 4378–4385, 2018.

[22] M. H. Segler, T. Kogej, C. Tyrchan, and M. P. Waller, "Generating focused molecule libraries for drug discovery with recurrent neural networks," *ACS Central Science*, vol. 4, no. 1, pp. 120–131, 2018.

[23] A. Gupta, A. T. Muller, B. J. Huisman, J. A. Fuchs, P. Schneider, and G. Schneider, "Generative recurrent networks for de novo drug design," *Molecular Informatics*, vol. 37, no. 1-2, p. 1700111, 2018.

[24] R. Gomez-Bombarelli, J. N. Wei, D. Duvenaud, J. M. Hernández-Lobato, B. Sanchez-Lengeling, D. Sheberla, J. Aguilera-Iparraguirre, T. D. Hirzel, R. P. Adams, and A. Aspuru-Guzik, "Automatic chemical design using a data-driven continuous representation of molecules," *ACS Central Science*, vol. 4, no. 2, pp. 268–276, 2018.

[25] D. P. Kingma and M. Welling, "Auto-encoding variational bayes," *arXiv preprint arXiv:1312.6114*, 2013.

[26] N. De Cao and T. Kipf, "De novo design of new chemical entities with reinvent," *ACS Central Science*, vol. 4, no. 11, pp. 1708–1719, 2018.

[27] O. Prykhodko, S. V. Johansson, P.-C. Kotsias, J. Arus-Pous, E. J. Bjerrum, O. Engkvist, and H. Chen, "A de novo molecular generation method using latent vector based generative adversarial network," *Journal of Cheminformatics*, vol. 11, no. 1, pp. 1–13, 2019.

[28] I. Goodfellow, J. Pouget-Abadie, M. Mirza, B. Xu, D. Warde-Farley, S. Ozair, A. Courville, and Y. Bengio, "Generative adversarial nets," *Advances in Neural Information Processing Systems*, vol. 27, pp. 2672–2680, 2014.

[29] K. Madhawa, K. Ishiguro, K. Nakago, and M. Abe, "Graphnvp: An invertible flow model for generating molecular graphs," *arXiv preprint arXiv:1905.11600*, 2019.

[30] C. Zang and F. Wang, "Mermaid: Metaphors for molecular generation," *arXiv preprint arXiv:2004.08047*, 2020.

[31] M. Olivecrona, T. Blaschke, O. Engkvist, and H. Chen, "Molecular denovo design through deep reinforcement learning," *Journal of Cheminformatics*, vol. 9, no. 1, pp. 1–14, 2017.

[32] M. Popova, O. Isayev, and A. Tropsha, "Deep reinforcement learning for de novo drug design," *Science Advances*, vol. 4, no. 7, p. eaap7885, 2018.

[33] S. Honda, S. Shi, and H. R. Ueda, "Smiles transformer: Pre-trained molecular fingerprint for low data drug discovery," *arXiv preprint arXiv:1911.04738*, 2019.

[34] Ł. Maziarka, A. Pocha, J. Kaczmarczyk, K. Rataj, and M. Warchoł, "Mol-cyclegan: a generative model for molecular optimization," *Journal of Cheminformatics*, vol. 12, no. 1, pp. 1–18, 2020.

[35] R. Irwin, S. Dimitriadis, J. He, and E. J. Bjerrum, "The molecular transformer for organic reaction prediction and symbolic inference," *Journal of Chemical Information and Modeling*, vol. 62, no. 13, pp. 3204–3213, 2022.

[36] K. Preuer, P. Renz, T. Unterthiner, S. Hochreiter, and G. Klambauer, "Frechet chemnet distance: A metric for generative models for molecules in drug discovery," *Journal of Chemical Information and Modeling*, vol. 58, no. 9, pp. 1736–1741, 2018.

[37] P. Ertl and A. Schuffenhauer, "Estimation of synthetic accessibility score of drug-like molecules based on molecular complexity and fragment contributions," *Journal of Cheminformatics*, vol. 1, no. 1, pp. 1–11, 2009.

[38] T. Sterling and J. J. Irwin, "Zinc 15–ligand discovery for everyone," *Journal of Chemical Information and Modeling*, vol. 55, no. 11, pp. 2324–2337, 2015.

[39] J. B. Baell and G. A. Holloway, "The pains problem in virtual screening," *Journal of Medicinal Chemistry*, vol. 53, no. 7, pp. 2719–2740, 2010.

[40] G. W. Bemis and M. A. Murcko, "The properties of known drugs. 1. molecular frameworks," *Journal of Medicinal Chemistry*, vol. 39, no. 15, pp. 2887–2893, 1996.

[41] W. Jin, R. Barzilay, and T. Jaakkola, "Junction tree variational autoencoder for molecular graph generation," *International Conference on Machine Learning*, pp. 2323–2332, 2018.

[42] S. A. Wildman and G. M. Crippen, "Prediction of physicochemical parameters by atomic contributions," *Journal of Chemical Information and Computer Sciences*, vol. 39, no. 5, pp. 868–873, 1999.

[43] G. R. Bickerton, G. V. Paolini, J. Besnard, S. Muresan, and A. L. Hopkins, "Quantifying the chemical beauty of drugs," *Nature Chemistry*, vol. 4, no. 2, pp. 90–98, 2012.

[44] J. Arus-Pous, T. Blaschke, S. Ulander, J.-L. Reymond, H. Chen, and O. Engkvist, "Randomized smiles strings improve the quality of molecular generative models," *Journal of Cheminformatics*, vol. 11, no. 1, pp. 1–13, 2019.

[45] R. Winter, F. Montanari, F. Noe, and D.-A. Clevert, "Learning con-´ tinuous and data-driven molecular descriptors by translating equivalent chemical representations," *Chemical Science*, vol. 10, no. 6, pp. 1692–1701, 2019.

[46] T. Blaschke, M. Olivecrona, O. Engkvist, J. Bajorath, and H. Chen, "Application of generative autoencoder in de novo molecular design," *Molecular Informatics*, vol. 37, no. 1-2, p. 1700123, 2018.

[47] M. Moret, L. Friedrich, F. Grisoni, D. Merk, and G. Schneider, "Generative models for molecular discovery: Recent advances and challenges," *Wiley Interdisciplinary Reviews: Computational Molecular Science*, vol. 10, no. 3, p. e1460, 2020.

[48] P. Maragakis, B. Alldritt, M. Ashton, S. Hennel, and J. B. Sorge, "Machine learning force fields and coarse-grained variables in molecular dynamics: application to materials and biological systems," *Journal of Chemical Theory and Computation*, vol. 16, no. 8, pp. 4651–4665, 2020.

[49] N. Yoshikawa, K. Terayama, M. Sumita, T. Homma, K. Oono, and K. Tsuda, "Population-based de novo molecule generation, using grammatical evolution," *Chemistry Letters*, vol. 47, no. 11, pp. 1431–1434, 2019.

[50] C. W. Coley, N. S. Eyke, and K. F. Jensen, "Autonomous discovery in the chemical sciences part i: Progress," *Angewandte Chemie International Edition*, vol. 59, no. 51, pp. 22858–22893, 2020.

[51] Y. Xu, K. Lin, S. Wang, L. Wang, C. Cai, C. Song, L. Lai, and J. Pei, "Deep learning for drug design: an artificial intelligence paradigm for drug discovery in the big data era," *Medicinal Research Reviews*, vol. 41, no. 4, pp. 1734–1766, 2021.

[52] J. Panteleev, H. Gao, and L. Jia, "Recent applications of machine learning in drug discovery," *Bioorganic & Medicinal Chemistry Letters*, vol. 28, no. 17, pp. 2807–2815, 2018.

[53] S. R. Atance, J. V. Diez, and O. Engkvist, "A practical guide to molecular generative models," *Journal of Cheminformatics*, vol. 14, no. 1, pp. 1–19, 2022.

[54] J. Jimenez-Luna, M. Skalic, N. Weskamp, and G. Schneider, "Exploration strategies for discovery of interacting multi-target drugs," *Drug Discovery Today*, vol. 25, no. 2, pp. 261–267, 2020.

[55] M. A. Sellwood, M. Ahmed, M. H. Segler, and N. Brown, "Artificial intelligence in drug discovery," *Future Medicinal Chemistry*, vol. 10, no. 17, pp. 2025–2028, 2018.

[56] M. Thomas, R. T. Smith, N. M. O'Boyle, C. de Graaf, and A. Bender, "Molgpt: Molecular generation using a transformer-decoder model," *Journal of Cheminformatics*, vol. 14, no. 1, pp. 1–19, 2022.

[57] G. Corso, H. Stark, B. Jing, R. Barzilay, and T. Jaakkola, "Diffdock:¨ Diffusion steps, twists, and turns for molecular docking," *arXiv preprint arXiv:2210.01776*, 2022.